\documentclass[aps,prd,showpacs,amsmath,10pt,twocolumn,superscriptaddress,floatfix,nofootinbib,notitlepage]{revtex4-2}
\usepackage[utf8]{inputenc}
\usepackage[english]{babel}
\usepackage[dvipsnames]{xcolor}
\usepackage[normalem]{ulem}
\usepackage{slashed}
\usepackage[
colorlinks=true,
linkcolor=blue,
breaklinks=true,
urlcolor=magenta,
citecolor=blue]{hyperref}

\usepackage{bm,color,xcolor,amsfonts,amsmath,amssymb,epsfig,esint,orcidlink}

\usepackage{soul}

\newcommand{\be}{\begin{equation}}
\newcommand{\ee}{\end{equation}}
\newcommand{\jp}{J/\psi}

\newcommand{\chisq}{\chi^2/{\rm dof}}
\newcommand{\mJ}{m_{J/\psi}}

\graphicspath{{Figs.dir/}}

\newcommand{\rub}{\affiliation{Institut f\"ur Theoretische Physik II, Ruhr-Universit\"at Bochum, D-44780 Bochum, Germany }}

\newcommand{\fzj}{\affiliation{Institute for Advanced Simulation, Forschungszentrum J\"ulich, D-52425 J\"ulich, Germany}}

\newcommand{\itp}{\affiliation{CAS Key Laboratory of Theoretical Physics, Institute of Theoretical Physics, \\Chinese Academy of Sciences, Beijing 100190, China}}

\newcommand{\ucas}{\affiliation{School of Physical Sciences, University of Chinese Academy of Sciences, Beijing 100049, China}}

\newcommand{\lis}{\affiliation{CeFEMA, Center of Physics and Engineering of Advanced Materials, Instituto Superior T{\'e}cnico, Avenida Rovisco Pais 1, 1049-001 Lisbon, Portugal}}

\newcommand{\peng}{\affiliation{Peng Huanwu collaborative Center for Research and Education, Beihang University, Beijing 100191, China}}

\newcommand{\scnt}{\affiliation{Southern Center for Nuclear-Science Theory (SCNT), Institute of Modern Physics,\\ Chinese Academy of Sciences, Huizhou 516000, China}}

\newcommand{\hiskp}{\affiliation{Helmholtz-Institut f\"ur Strahlen- und Kernphysik% %and Bethe Center for Theoretical Physics
, Universit\"at Bonn, 53115 Bonn, Germany}}

\newcommand{\hohai}{\affiliation{College of Mechanics and Engineering Science, Hohai University, Nanjing 211100, China}}

\definecolor{darkgreen}{rgb}{0.0, 0.6, 0.0}

\begin{document}

\title{Towards a precision determination of the $X(6200)$ parameters from data}

\author{Yi-Lin~Song\orcidlink{0009-0005-9302-1349}}\email{songyilin@itp.ac.cn}
\itp \ucas 

\author{Yu Zhang\orcidlink{0009-0007-2951-6536}}\email{2110020124@hhu.edu.cn}
\hohai \itp

\author{Vadim Baru\orcidlink{0000-0001-6472-1008}}\email{vadim.baru@tp2.rub.de}
\rub

\author{Feng-Kun~Guo\orcidlink{0000-0002-2919-2064}}\email{fkguo@itp.ac.cn}
\itp \ucas \peng \scnt

\author{Christoph~Hanhart\orcidlink{0000-0002-3509-2473}}\email{c.hanhart@fz-juelich.de}
\fzj

\author{Alexey Nefediev\orcidlink{0000-0002-9988-9430}}\email{a.nefediev@uni-bonn.de}
\hiskp
\lis

\begin{abstract}
In a recent paper, Phys. Rev. Lett. {\bf 126}, 132001 (2021), the LHCb data on the di-$J/\psi$ production in proton-proton collisions were analysed in a coupled-channel framework based on double-vector-charmonium channels. 
This investigation identified a robust pole near the $J/\psi J/\psi$ threshold, tagged $X(6200)$, suggesting it as a new state. The present work extends that investigation by incorporating recent di-$J/\psi$ production data 
from the CMS and ATLAS Collaborations and
performing a combined analysis of all three datasets.
This study confirms the existence of the $X(6200)$, and its pole position is now determined with a higher precision than in the previous study, where only a single dataset was employed.
The pole corresponding to the $X(6900)$ is also extracted, though its actual position and residue depend on a particular coupled-channel model employed and deviate from values reported in the experimental investigations.
Moreover, we demonstrate
that the currently available data do not allow one to determine whether there is an additional pole in the studied mass range.
The di-electron width of the $X(6200)$ is estimated under different conjectures on its quantum numbers, and the number of such states that can be annually produced at the future Super $\tau$-Charm Facility is estimated. The line shape in the complimentary $J/\psi\psi(2S)$ channel is discussed, and a good agreement with the ATLAS data is found.
\end{abstract}

\maketitle

\section{Introduction}

The last two decades are marked with considerable progress in experimental studies of the spectrum of hadrons containing heavy quarks. The properties of many discovered states cannot be understood in the framework of the most simple quark model that had previously been regarded as a single unified classification scheme for all hadrons. For recent reviews of such unconventional, or exotic, states and various approaches suggested for their interpretation see Refs.~\cite{Hosaka:2016pey,Lebed:2016hpi,Esposito:2016noz,Guo:2017jvc,Olsen:2017bmm,Liu:2019zoy,Brambilla:2019esw,Guo:2019twa,Yang:2020atz,Liu:2024uxn, Husken:2024rdk, Zhu:2024swp, ParticleDataGroup:2024cfk,Chen:2024eaq}.

While most of experimental information arrives for hadrons containing two heavy constituents, the LHCb measurements of the di-$J/\psi$ production in proton-proton collisions at the centre-of-mass (c.m.) energies 7, 8, and 13~TeV revealed a new potentially rich class of exotic states with four charmed (anti)quarks~\cite{Aaij:2020fnh}. Indeed, the measured line shape has a nontrivial form, departing
significantly from the phase space distribution as well as the exponential behaviour from single and double-parton scatterings. Attention of both experimental and theoretical communities was mainly attracted to the statistically significant peaking structures in the signal observed in the energy range from the double-$\jp$ threshold at 6.2~GeV to approximately 7.2~GeV and, in particular, the most pronounced structure was named $T_{cc\bar{c}\bar{c}}(6900)$ [also known as $X(6900)$] \cite{ParticleDataGroup:2024cfk}.
However, when the data are analysed using a coupled-channel scattering formalism with vector-charmonium pairs~\cite{Dong:2020nwy},
the number and locations of poles in the 6900 MeV mass region appear
rather badly determined. Nonetheless,
if this approach indeed captures
the relevant dynamics,
the data suggest the existence of a pole near the di-$J/\psi$ threshold in the double-$J/\psi$ production amplitude. 
This state was named $X(6200)$ (or $T_{cc\bar{c}\bar{c}}(6200)$ according to the new naming scheme for exotica promoted by the Particle Data Group~\cite{ParticleDataGroup:2024cfk}). The finding was confirmed by the calculation in Ref.~\cite{Liang:2021fzr} and more recently also in Ref.~\cite{Huang:2024jin}. 
Given the suppression of the signal near the threshold caused by the phase space factor, this pole cannot reveal itself as a pronounced peaking structure above the threshold in the double-$J/\psi$ line shape. Instead, it can only be unambiguously identified through a comprehensive pole search in
coupled-channel scattering amplitudes.
Meanwhile, a steep rise of the line shape just above the threshold provides evidence for the existence of such a near-threshold pole~\cite{Guo:2014iya}. As follows from the analysis in Ref.~\cite{Dong:2020nwy}, the behaviour of the signal just above the di-$J/\psi$ threshold indeed calls for the existence of the $X(6200)$ pole. 

Recently, the data on the double-$J/\psi$ production in $pp$ collisions also arrived from two other LHC Collaborations, namely CMS~\cite{CMS:2023owd} and ATLAS~\cite{ATLAS:2023bft}. Remarkably, in both cases, the data description improved after the inclusion of an auxiliary Breit-Wigner resonance centred just above the production threshold. The necessity of its inclusion to the fitting function supports the existence of the $X(6200)$ pole in the amplitude.
Thus, in this paper, we employ the models previously developed in Ref.~\cite{Dong:2020nwy} to perform a combined analysis of all datasets from LHCb, CMS, and ATLAS, to check whether the same near-threshold pole is consistent with all three datasets. As a result of this analysis, we confirm the existence of the $X(6200)$ pole and more precisely extract its position on the energy complex plane, while putting further doubts to the resonance parameters extracted in the well-above-threshold regime.

\begin{table*}[t!]
\caption{Channels with the thresholds lying within the energy range 6.2--7.1~GeV. The tags ``2 ch.'' and ``3 ch.'' indicate the models that include the tagged channel; all disregarded channels are marked with ``\ldots'' (see the main text for details).}
\resizebox{\textwidth}{!}{
\begin{ruledtabular}
\begin{tabular}{cccccccccc}
Channel& $J/\psi J/\psi$&
$J/\psi h_c$ &
$J/\psi \psi(2S)$ &
$\chi_{c0}\chi_{c0}$ &
$\chi_{c0}\chi_{c1}$ &
$J/\psi \psi(3770)$ &
$\chi_{c1}\chi_{c1}$ &
$h_ch_c$ &
$\chi_{c2}\chi_{c2}$ \\
%$\psi(2S)h_c$ \\
\hline
Threshold& 6.194 &
6.622 &
6.783 &
6.830 &
6.925 &
6.870 &
7.021 &
7.051 &
7.132 \\
%7.211 \\
\hline
Model&2 and 3 ch.& \ldots &2 and 3 ch.& \ldots & \ldots &3 ch.& \ldots & \ldots & \ldots
\end{tabular}
\end{ruledtabular}
}
\label{tab:chan}
\end{table*}

\begin{figure}[t]
\centering
\includegraphics[width=\linewidth]{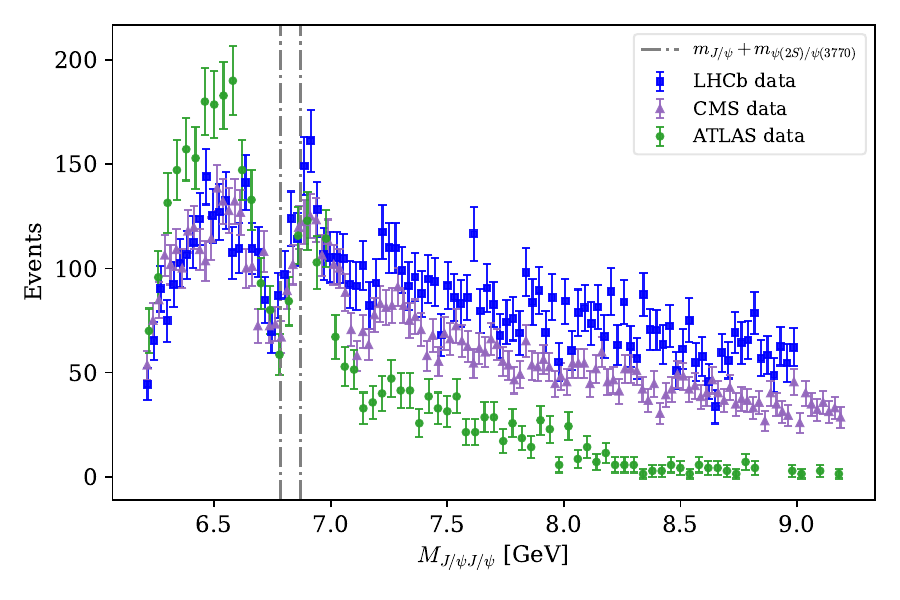}
\caption{The data on di-$J/\psi$ production in proton-proton collisions from the LHCb \cite{Aaij:2020fnh}, CMS \cite{CMS:2023owd}, and ATLAS \cite{ATLAS:2023bft} Collaborations.
For presentation purposes, the CMS and ATLAS data have been rescaled by factors of $25/28$ and $40/28$, respectively, to normalise them to the LHCb data.
The $J/\psi\psi(2S)$ and $J/\psi\psi(3770)$ thresholds are shown as vertical dashed lines.}
\label{fig:data}
\end{figure}

\section{The coupled-channel models}
\label{sec:model}

\begin{table*}[tbh]
\caption{Parameters of the combined two-channel fit. Parameters with bars quoted here differ from the parameters without bars introduced in Eq.~\eqref{eq:V2c} as explained in Eq.~\eqref{eq:bars}.}
\label{tab:params2c}
\resizebox{\textwidth}{!}{
\begin{ruledtabular}
 \begin{tabular}{ccccccccccccc}
 $\bar{a}_1$ [GeV$^{-2}$] & $\bar{a}_2$ [GeV$^{-2}$] & $\bar{b}_1$ [GeV$^{-4}$] & $\bar{b}_2$ [GeV$^{-4}$]& $\bar{c}$ [GeV$^{-2}$]& $\alpha_1$ & $\alpha_2$ & $\alpha_3$ & $b$ & $\chisq$\\
 \hline
 $0.32^{+0.31}_{-0.29}$ & $-4.11^{+0.34}_{-0.33}$ & $-1.74_{-0.25}^{+0.22}$ & $-6.50_{-0.40}^{+0.40}$ & $3.03^{+0.23}_{-0.26}$ &
 $79^{+6}_{-5}$ & $108^{+9}_{-8}$ & $560_{-40}^{+40}$ & $2.89^{+0.23}_{-0.22}$& 1.32\\
 \end{tabular}
 \end{ruledtabular}
 }
 \caption{The same as in Table~\ref{tab:params2c} but for the three-channel fit. All parameters $\bar{a}_{ij}$ are given in the units of GeV$^{-2}$.}
 \label{tab:params3c}
 \resizebox{\textwidth}{!}{
 \begin{ruledtabular}
 \begin{tabular}{cccccccccccccc}
 &$\bar a_{11}$ &$\bar a_{12}$&$\bar a_{13}$ &$\bar a_{22}$ &$\bar a_{23}$ &$\bar a_{33}$&$\alpha_1$ & $\alpha_2$ & $\alpha_3$ & $b$ & $\chisq$\\
 \hline
 fit 1 &$6.5^{+1.4}_{-1.4}$ & $10.4^{+2.5}_{-2.3}$ & $-0.8_{-1.0}^{+1.7}$ & $12^{+4}_{-4}$ & $-3.0^{+2.0}_{-0.9}$ &$-1.7^{+1.0}_{-1.0}$ &$308^{+84}_{-91}$ &$425^{+116}_{-126}$ & $2185^{+596}_{-421}$ & $0.09^{+0.15}_{-0.21}$ & 1.32\\
 fit 2 &$8.9^{+2.1}_{-1.5}$ & $17.8^{+3.3}_{-2.4}$ & $2.8^{+1.7}_{-1.7}$ & $29^{+5}_{-4}$ & $0.5^{+3.1}_{-3.1}$ & $-2.5^{+0.5}_{-0.4}$ & $164^{+21}_{-16}$ & $226^{+29}_{-23}$ & $1161^{+151}_{-116}$ & $-0.50^{+0.17}_{-0.16}$& 1.40\\
 \end{tabular}
 \end{ruledtabular}
 }
\end{table*}

In this section we briefly introduce the minimal coupled-channel models suggested in Ref.~\cite{Dong:2020nwy} and discuss their modifications necessary to adapt them to analyse the datasets from different experimental Collaborations.
Since thresholds strongly affect line shapes, 
it is crucial to employ in the data analysis a coupled-channel approach that incorporates all relevant channels with the thresholds residing within the studied energy range (we choose it to be 6.2--7.1~GeV). On the other hand, to reduce the number of fitting parameters and in this way fit them reliably, only the most relevant channels need to be retained. We first enumerate all channels with the thresholds lying in the appropriate energy range and collect them in Table~\ref{tab:chan}. Then we disregard unimportant channels~\cite{Dong:2020nwy}, namely those that involve $h_c$ as suppressed in the heavy quark limit by heavy quark spin symmetry (HQSS) and
$\chi_{cJ}\chi_{cJ'}$ ones, with $J,J'=0,1,2$, as suppressed by exchanging heavier light mesons (with the $\omega$ exchange being the lightest allowed one) than that between two $\psi$ states (two pions can be exchanged as hadronisation of the soft gluons~\cite{Dong:2021lkh}). We, therefore, arrive at two coupled-channel models with the following sets of the included channels:
\begin{itemize}
\item \{$J/\psi J/\psi,J/\psi\psi(2S)$\} referred to as the two-channel model;
\item \{$J/\psi J/\psi,J/\psi\psi(2S),J/\psi\psi(3770)$\} referred to as the three-channel model.
\end{itemize}

In Fig.~\ref{fig:data}, we show all three analysed datasets together and conclude that they demonstrate very similar peaking structures in the energy range of interest 6.2--7.1~GeV.
One can see that the positions of the sharpest structures in the measured line shapes indeed coincide with the thresholds of the channels selected above, so the choice of these two coupled-channel schemes is justified.
It is worthwhile noticing that both peak or dip can appear near threshold in the line shape, depending on the interference pattern~\cite{Dong:2020hxe, Zhang:2024qkg}. 

The interaction potentials for the two- and three-channel model are taken in the form
\begin{equation}
V_{\rm 2ch}(E)=\begin{pmatrix}
a_1 + b_1 k_1^2 & c \\
 c & a_2 + b_2 k_2^2
\end{pmatrix},
\label{eq:V2c}
\end{equation}
and
\begin{equation}
V_{\rm 3ch}(E)=\begin{pmatrix}
a_{11} & a_{12} & a_{13} \\
a_{12} & a_{22} & a_{23} \\
a_{13} & a_{23} & a_{33}
\end{pmatrix},
\label{eq:V3c}
\end{equation}
respectively, where $\{a_1,a_2,b_1,b_2,c\}$ and $\{a_{ij}\}$ ($i,j=1,2,3$) are real free parameters, $E$ is the c.m. energy, and $k_i=\lambda^{1/2}(E^2,m_{i1}^2,m_{i2}^2)/(2E)$ is the magnitude of the corresponding three-momentum in the $i$th channel with $\lambda(x,y,z)=x^2+y^2+z^2 - 2xy - 2yz - 2xz$ for the K\"all\'en triangle function.
In principle, we could also allow for energy dependencies in terms other than the diagonals of the two-channel model, however, the parameterisations shown above already lead to a satisfactory description of the data such that additional parameters could not be constrained empirically.

Since the $T$ matrix entering the final state interaction is universal, the $T$-matrix parameters in Eqs.~\eqref{eq:V2c} and \eqref{eq:V3c} for the two different models should not depend on the details of the experiment and as such can be fixed from a combined fit to all the data in Fig.~\ref{fig:data}.
To stick to the same notation as in Ref.~\cite{Dong:2020nwy}, we also define nonrelativistic parameters of the potentials (tagged with bars) defined through the relation 
\be
V_{ij}(E)=\sqrt{2m_i\; 2m_j}\;\bar{V}_{ij}(E),
\label{eq:bars}
\ee
with $m_i$ for the mass of the corresponding charmonium state, $\mJ$, $m_{\psi(2S)}$, or $m_{\psi(3770)}$. That is, we fit the data with the parameters $\{\bar{a}_1,\bar{a}_2,\bar{b}_1,\bar{b}_2,\bar{c}\}$ for the two-channel model and $\{\bar{a}_{ij}\}$ for the three-channel model, with the parameters in Eqs.~\eqref{eq:V2c} and \eqref{eq:V3c} obtained from the barred ones through multiplying them by $\sqrt{2m_i\; 2m_j}$.

In our notation, the $S$-wave $S$ and $T$ matrices are related as
\be
S_{ij}(E) = 1 - 2i \sqrt{\rho_i(E)\rho_j(E)} T_{ij}(E),\quad \rho_i(E) = \frac{k_i}{8\pi E},
\ee
where the scattering $T$ matrix comes as a solution to the Lippmann-Schwinger equation,
\be
T(E)=V(E)\cdot[1-G(E)\cdot V(E)]^{-1},
\label{eq:T}
\ee
with $V(E)$ for the relevant potential matrix in Eq.~\eqref{eq:V2c} or \eqref{eq:V3c} above and $G(E)={\rm diag}(G_1,G_2,\ldots)$ for the diagonal matrix composed of the propagators in the corresponding two-body channels,
\begin{align}
&G_i(E)=i\int \frac{d^4q}{(2\pi)^4}\frac1{(q^2-m_{i1}^2+i\epsilon)[(P-q)^2-m_{i2}^2+i\epsilon]}\nonumber\\
&=\frac1{16\pi^2}\bigg\{a(\mu)+\log\frac{m_{i1}^2}{\mu^2}+\frac{m_{i2}^2-m_{i1}^2+s}{2s} \log\frac{m_{i2}^2}{m_{i1}^2} \nonumber\\
&~~~+\frac{k_i}{E} \Big[
\log\left(2k_i E+s+\Delta_i\right) +
\log\left(2k_i E+s-\Delta_i\right) \nonumber\\
&~~~ -
\log\left(2k_i E-s+\Delta_i\right) -
\log\left(2k_i E-s-\Delta_i\right)
\Big]\bigg\},\label{eq:GDR}
\end{align}
where $s={P^2}=E^2$, $m_{i1}$ and $m_{i2}$ are the particle masses in the $i$th channel, and $\Delta_i= m_{i1}^2-m_{i2}^2$. The parameter $\mu$ denotes the dimensional regularisation scale, and $a(\mu)$ is a subtraction constant. We set $\mu=1$ GeV, and $a(\mu = 1\text{ GeV}) = -3$ as in Ref.~\cite{Dong:2020nwy}; its variation in Eq.~\eqref{eq:T} can be absorbed into a change of the contact term parameters in the potential. We emphasise that the $T$ matrix in Eq.~\eqref{eq:T} is fully fixed,
once the parameters are fitted to the data. 

\begin{figure*}[tbhp]
\centering
\includegraphics[width=0.45\linewidth]{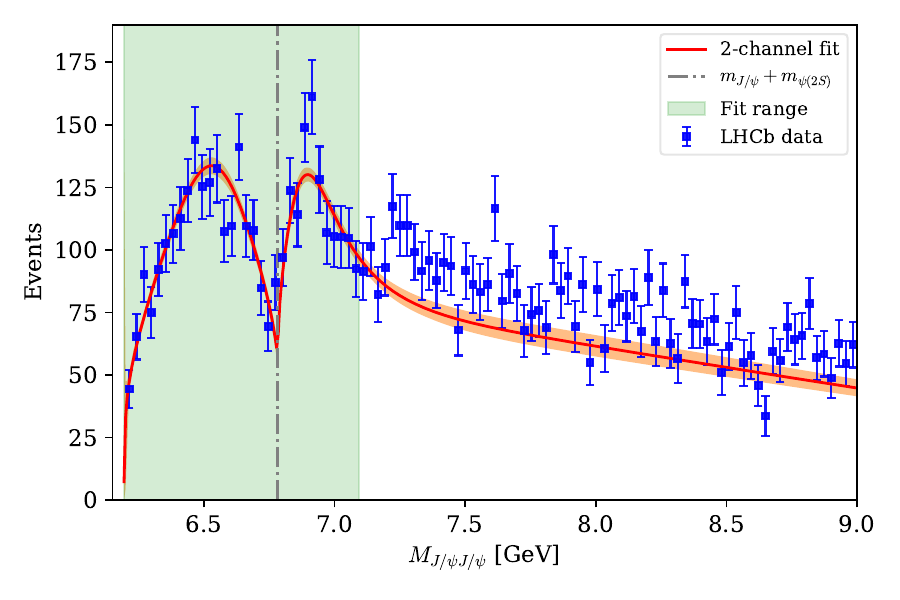}
\includegraphics[width=0.45\linewidth]{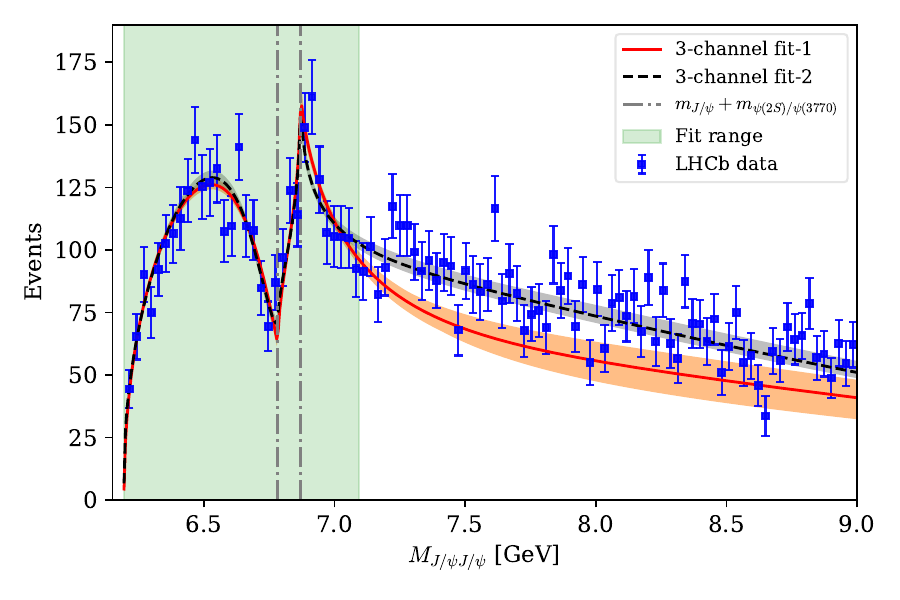}\\
\includegraphics[width=0.45\linewidth]{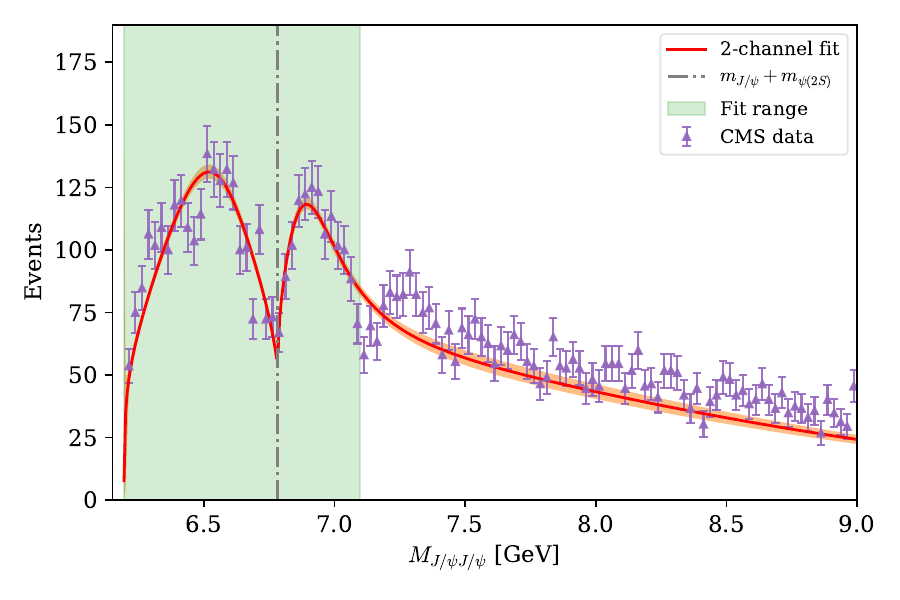}
\includegraphics[width=0.45\linewidth]{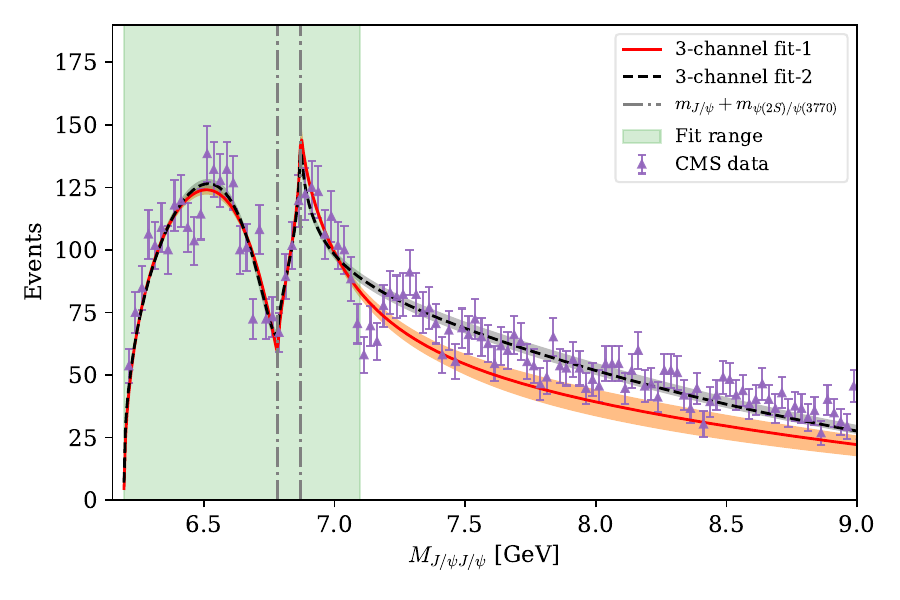}\\
\includegraphics[width=0.45\linewidth]{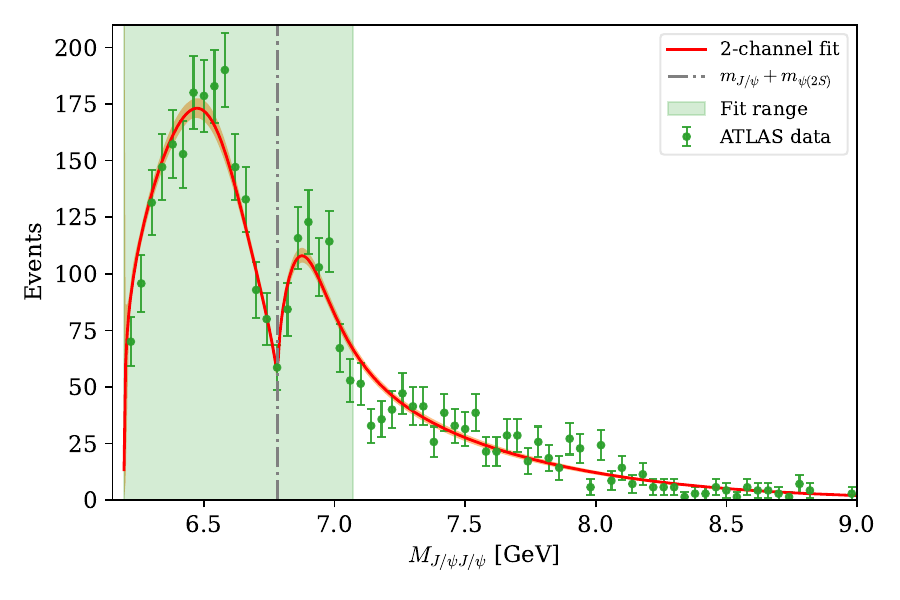}
\includegraphics[width=0.45\linewidth]{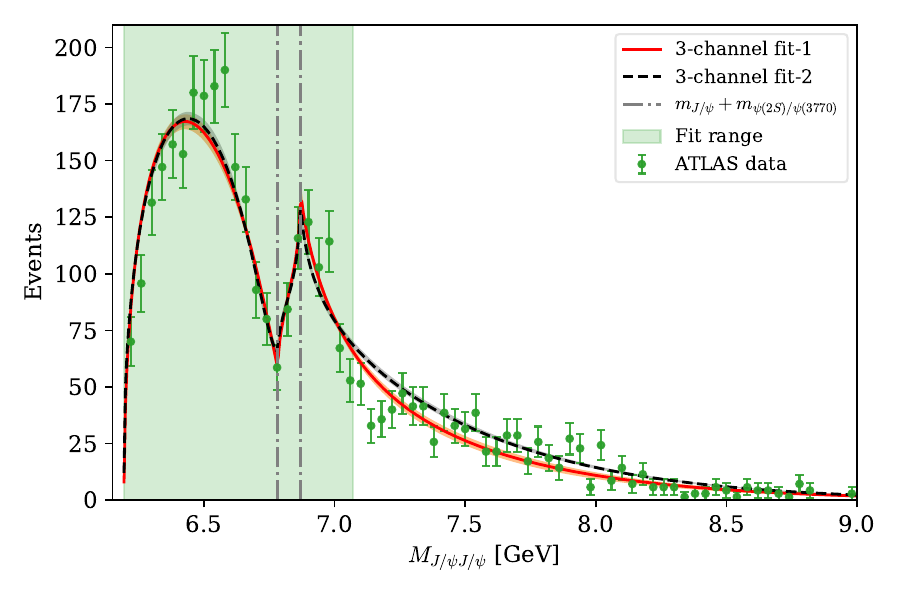}
\caption{Combined fits to the LHCb (first row), CMS (second row), and ATLAS (third row) data. The results for the two-channel (1 fit) and three-channel (2 fits of similar quality) models are shown in the left and right panels, respectively. The dashed vertical lines indicate the position of the relevant thresholds taken into account in addition to the production channel $J/\psi J/\psi$, that is, $J/\psi\psi(2S)$ and $J/\psi\psi(3770)$; see Table~\ref{tab:chan} and the discussion in the main text. The fitted energy range is shaded with light green.}
\label{fig:fits}
\end{figure*}
 
The production amplitude in the channel $J/\psi J/\psi$ is built as
\be
\mathcal{M}_1 = P(E)\left (b + G_1T_{11}+\sum_{i = 2,3}r_iG_iT_{1i}\right),
\ee
where $P(E) = \alpha e^{-\beta E^2}$ is the short-distance source function producing the double-$J/\psi$ system from parton-level scattering that can be fixed from the large-energy tail of the experimental line shape. It is clearly seen from Fig.~\ref{fig:data} that the three datasets demonstrate different patterns above approximately 7.1~GeV that must be attributed to different, experiment-specific production mechanisms. To account for this fact, we employ 3 different sets of the background parameters, $\{\alpha_i,\beta_i\}$ with $i=1,2,3$ for LHCb, CMS, and ATLAS, respectively, and use them accordingly when building the individual contributions from different datasets to the overall $\chi^2$ in the combined fits. Then the $\beta_i$ parameters (in GeV$^{-2}$),
\be
\beta_1=0.012,\quad 
\beta_2=0.020,\quad
\beta_3=0.056,
\ee
are fixed by fitting the high-energy tails of the corresponding line shapes, which should be due to parton-level scattering, while the normalisation parameters, $\alpha_i$, are treated as free ones in the fit. 
The ratios $r_2$ and $r_3$ mimic potentially different production mechanisms for different channels. They are set to 1 as the fits do not call for different values ($r_3=0$ for the two-channel fit).
Following the logic of Ref.~\cite{Dong:2020nwy}, we introduce the parameter $b\neq 1$ that accounts for a possible violation of unitarity in the inclusive production amplitude, which may be present in a 2-body treatment given the complexity of the inclusive reaction from which the data were extracted (for example, from rescattering of $J/\psi$ with spectator particles that are not detected).

\section{Results and discussion}

\begin{table*}[tbh]
\caption{Pole positions (in MeV) for the models employed in the analysis. The Riemann sheets (RSs) quoted in parentheses are tagged with the signs of the imaginary parts of the momenta in all channels relevant for the particular model.}
\label{tab:poles}
\resizebox{\textwidth}{!}{
\begin{ruledtabular}
\begin{tabular}{cccc}
Model & Two-channel fit & Three-channel fit~1 & Three-channel fit~2\\
\hline
$X(6200)$ &$[6171,6194]$ (RS$_{++}$)
or $[6124,6202]-i[0,12]$ (RS$_{-+}$)
&$6151_{-16}^{+12}$ (RS$_{+++}$)&$6189_{-6}^{+3}$ (RS$_{-++}$)\\
$X(6900)$ &$6831_{-17}^{+15} - i162_{-12}^{+8}$ (RS$_{--}$)&$7071_{-101}^{+107} - i147_{-55}^{+81}$ (RS$_{-++}$)&$6872_{-45}^{+77} - i148_{-43}^{+37}$ (RS$_{-++}$)\\
Other poles &$6564_{-14}^{+13}-i282_{-27}^{+27}$ (RS$_{-+}$)& \ldots & \ldots
 \end{tabular}
 \end{ruledtabular}
 }
\end{table*}
 
\begin{figure*}[tbh]
\centering
\includegraphics[width=0.51\linewidth]{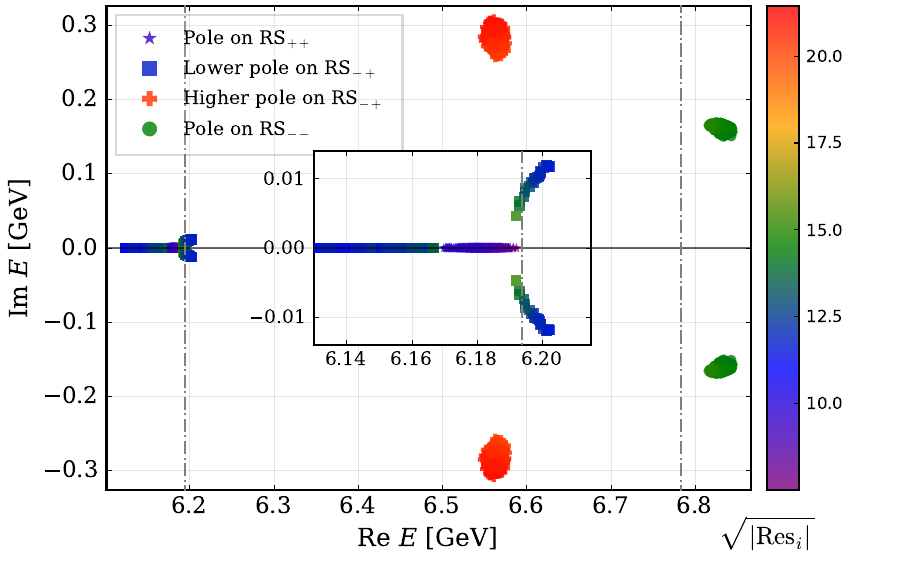}
\includegraphics[width=0.47\linewidth]{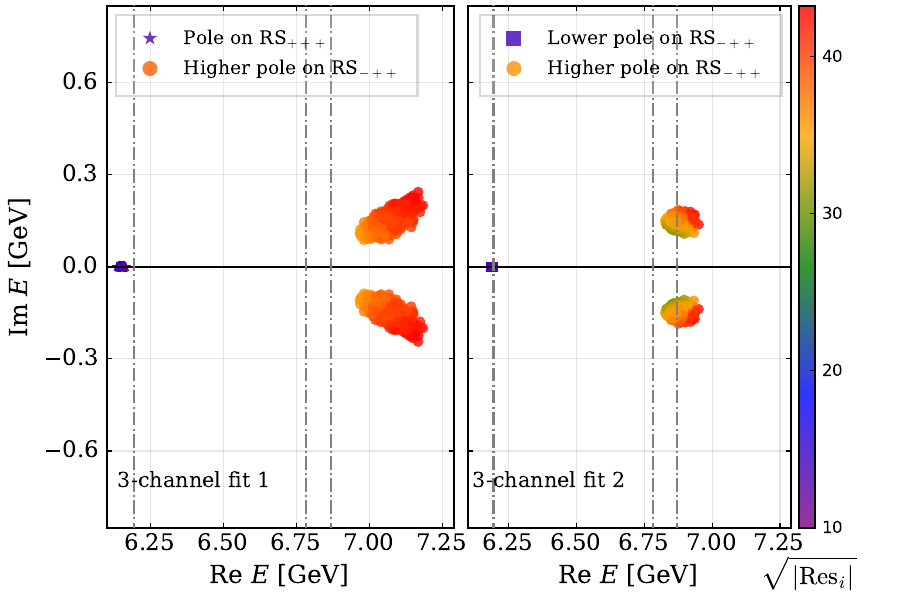
}
\caption{Pole positions for the two-channel (left) and three-channel (right) models. The colour indicates the value of the residue at the corresponding pole. In the inset subplot in the left panel, the pole trajectory is in fact continuous; the discontinuity therein is because the parameter region, where the conjugate pair of RS$_{--}$ poles approach the real axis, is very small and thus not covered completely by the random sampling of the $1\sigma$ parameter sets. }
\label{fig:poles}
\end{figure*}

In Fig.~\ref{fig:fits}, we show the results for the combined fits to the data in Fig.~\ref{fig:data} performed using the two-channel (left panel) and three-channel (right panel) models. The fitted values of all the parameters are listed in Tables~\ref{tab:params2c} and \ref{tab:params3c}.
Both models provide a decent description of the data of a similar quality, so they both can be employed for a
study of the pole content underlying the data. The extracted poles enumerated in Table~\ref{tab:poles} (see also Fig.~\ref{fig:poles}) confirm the pattern previously deduced from the analysis performed in Ref.~\cite{Dong:2020nwy} for the LHCb data alone, with much smaller uncertainties propagated from the data. Namely, all fits obtained using both models predict the existence of a pole near the double-$J/\psi$ threshold named the $X(6200)$. 
Depending on the parameters, it can reveal itself either as a bound-state pole on RS$_{++(+)}$ or as a near-threshold pole on RS$_{-+(+)}$.\footnote{The subscripts $\pm$ correspond to the signs of the imaginary parts of the respective coupled-channel momenta in the two(three)-channel case. In particular, bound or virtual state poles are located on the real axis of the RS$_{++(+)}$ and RS$_{-+(+)}$ sheet, respectively.}
We note that the appearance of ``or'' in both the pole location (see Table~\ref{tab:poles}) and the results for the scattering length (see Table~\ref{tab:a0r0XA}) does not imply fits with non-overlapping parameter regions. Instead, it is driven by the underlying physics since
the attraction strength in the double-$J/\psi$ system depends on the parameters that continuously change in the parameter space of the two-channel fit. If the attraction is strong enough, we have a bound-state pole on the Riemann sheet RS$_{++}$.
Then, for a decreased interaction strength, the bound-state pole approaches the double-$J/\psi$ threshold to move to RS$_{-+}$ and thus turn to a virtual state (the scattering length changes its sign passing through $\pm\infty$). Decreasing the interaction strength even further, the pole travels across RS$_{-+}$ to leave the real energy axis below the threshold, acquire an imaginary part, and eventually move above the threshold. A thorough discussion of the pole trajectories for a two-channel system can be found in Ref.~\cite{Zhang:2024qkg}.

In addition, we extracted the pole position that can be associated with the fully charmed tetraquark $X(6900)$.
Its mass (the real part of the pole position) from the two-channel fit lies sizably lower than the value averaged by the Particle Data Group, $(6899 \pm 12)$~MeV~\cite{ParticleDataGroup:2024cfk}. We note that the CMS Collaboration reported also a third structure at $7134^{+48}_{-25}\mbox{(stat)}^{+41}_{-15}\mbox{(syst)}~\mbox{MeV}$ \cite{CMS:2023owd} that, however, lies beyond the energy range analysed in this work. Since the values reported by the experimental groups were obtained using the Breit-Wigner parameterisation, the pinpointed difference indicates the importance of accounting for nearby threshold [$J/\psi\psi(2S)$ here] effects in the analysis of the line shape.
For the three-channel fits, analogous to Ref.~\cite{Dong:2020nwy}, the pole close to 6.9~GeV, that is near the third [$J/\psi \psi(3770)$] threshold, is located on a remote RS, RS$_{-++}$, whose path to the physical one needs to circle around the $J/\psi\psi(2S)$ threshold.
Then the pole shows up as a threshold cusp~\cite{Zhang:2024qkg} and its position does not correspond to a peak or dip location. 

Finally, a comment on the other poles is in order here. As follows from Table~\ref{tab:poles}, there is an additional pole with a real part around 6564~MeV on RS$_{-+}$ found in the two-channel fit. One may be tempted to assign this pole to the CMS and ATLAS $X(6600)$ state, which was introduced via a Breit-Wigner parameterisation. We notice, however, that
this pole does not appear in our three-channel fits.
The difference between the 2- and three-channel models
stems from different sources of the energy dependence in the potential necessary to produce a peak around 6.9~GeV. Indeed, in the former case, this energy dependence is explicitly provided by the $k_1^2$ and $k_2^2$ terms in Eq.~\eqref{eq:V2c} while, in the latter case, it implicitly comes from the coupling to the $J/\psi\psi(3770)$ channel. Consequently, the resulting $T$ matrices do not have the same pole contents. Therefore, our findings imply that the inference of the pole around 6.6~GeV from the available data is model dependent and not strictly necessary to reproduce the current data. This issue deserves a further investigation for an updated experimental input in the future.

Finally, we employ the effective range expansion of the scattering amplitude in the $\jp\jp$ channel,
\begin{equation}
T(k) = - 8\pi \sqrt{s} \left[\frac1{a_0} + \frac12 r_0 k^2 -i\, k + \mathcal{O}(k^4) \right]^{-1},
\end{equation}
to extract the $S$-wave scattering length $a_0$ and the effective range $r_0$. These parameters allow us to estimate the compositeness of the $X(6200)$~\cite{Matuschek:2020gqe},
\begin{equation}
\bar X_A =(1+2|r_0/a_0|)^{-1/2}.
\label{eq:XA}
\end{equation}
The results are presented in Table~\ref{tab:a0r0XA}, where only statistical uncertainties propagated from the data are provided. Notice that a large positive uncertainty $+0.62$ found for $\bar{X}_A$ in the two-channel fit is due to the fact that the parameter space covers a region for $a_0^{-1}$ very close to zero, cf. Eq.~\eqref{eq:XA}. An $\bar X_A$ value close to unity means a rather large double-$J/\psi$ composite component in the $X(6200)$ wave function. Thus our present results confirm the conclusions previously made in Ref.~\cite{Dong:2020hxe} that the molecular interpretation of the $X(6200)$ is preferred; however the two-channel model is also consistent with a sizable admixture of the compact component in the wave function [see, for example, Ref.~\cite{Nefediev:2021pww} for the discussion of a possible interplay between the quark and meson degrees of freedom (d.o.f.) in the $X(6200)$].
Studies of the final state consisting of two lepton pairs from the decays of the two $J/\psi$ particles with $Q^2\neq m_{J/\psi}^2$ should allow for a
more direct observation of the $X(6200)$ and also shed further light on its quantum numbers and nature. In particular, for a sufficiently fine resolution, a distinction between a bound state, virtual state,
and resonance should be straightforward.

\begin{table}[tbh]
\caption{The effective range parameters in the $J/\psi J/\psi$ channel and the compositeness $\bar X_A$ of $X(6200)$. The sign of the scattering length by convention is negative (positive) if $X(6200)$ is a bound (virtual) state.}
\label{tab:a0r0XA}
\begin{ruledtabular}
\begin{tabular}{cccc}
 &  Two-channel & Three-channel& Three-channel\\[-1mm]
 & fit & fit~1 & fit~2 \\
 \hline
 $a_0$(fm) &$\leq -0.31$
 or $\geq 0.78$
 &$-0.52_{-0.08}^{+0.07}$ &$1.63_{-0.38}^{+0.75}$ \\
 %\hline
 $r_0$(fm)&$-1.75_{-0.48}^{+0.16} $&$-0.05_{-0.01}^{+0.01}$ &$-0.08_{-0.02}^{+0.02}$ \\
 %\hline
 $\bar X_A$ &$0.37_{-0.09}^{+0.62}$ & $0.92_{-0.02}^{+0.02}$ & $0.95_{-0.02}^{+0.02}$
 \end{tabular}
 \end{ruledtabular}
\end{table}

\section{Dielectron width of $X(6200)$ and its direct production in $e^+e^-$ annihilation}

\begin{figure*}[htbp]
\includegraphics[width=0.3\linewidth]{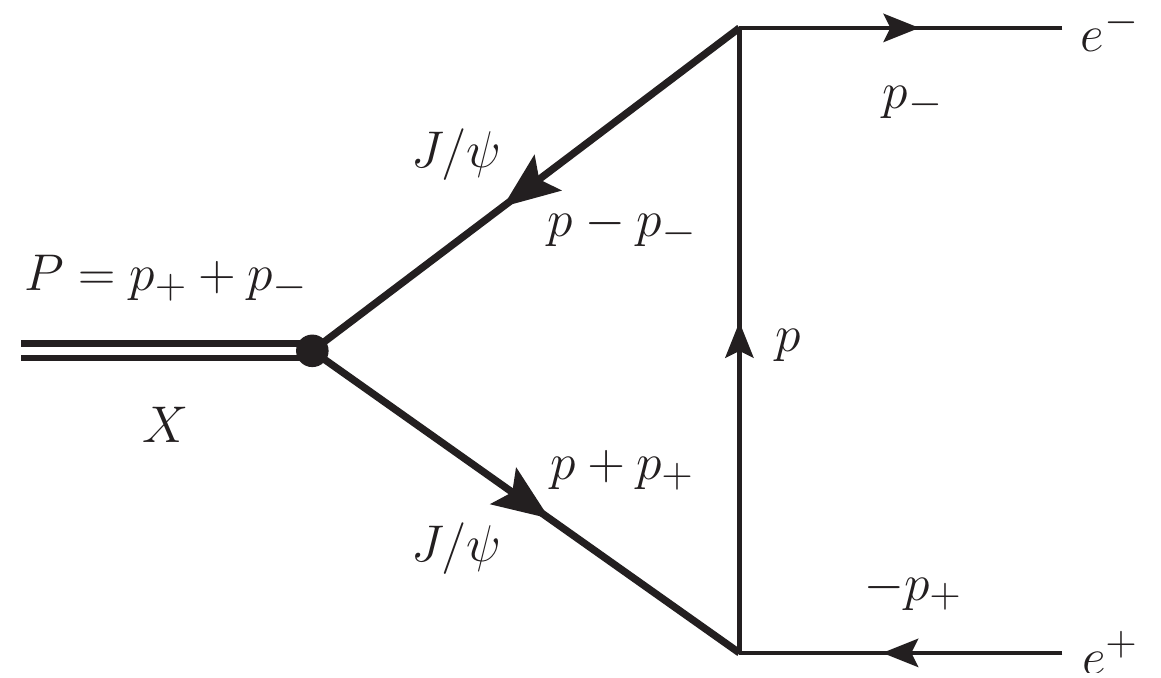}
\hspace*{0.05\linewidth}
\includegraphics[width=0.3\linewidth]{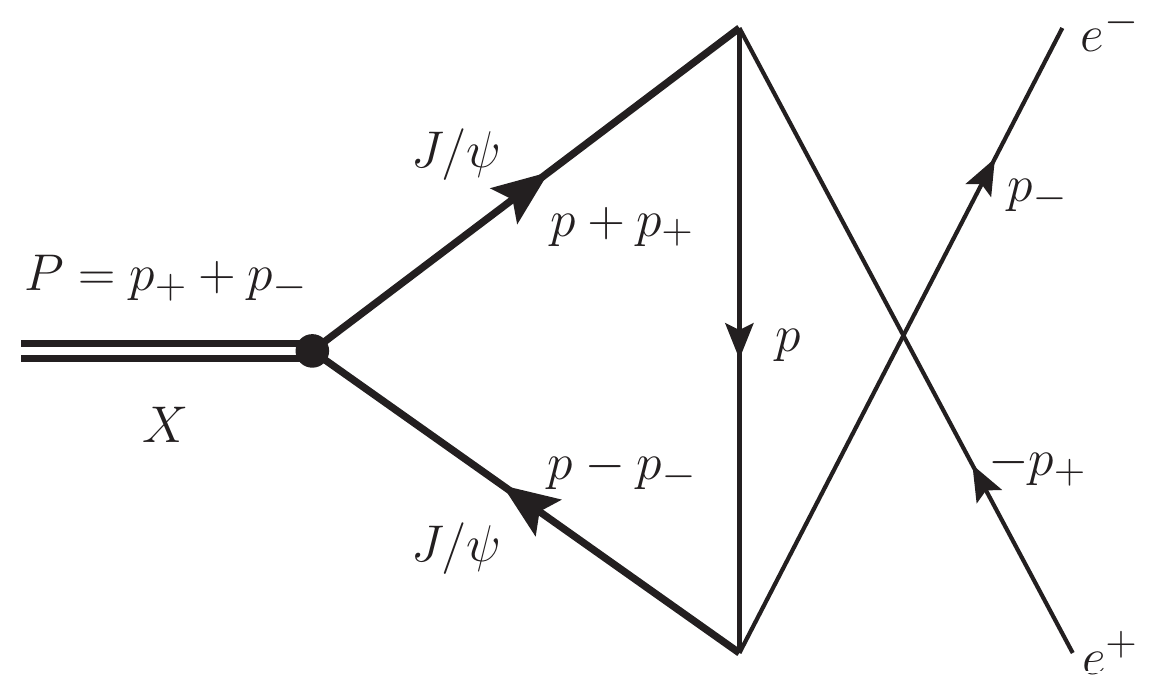}
\caption{The diagrams for the $X(6200)$ dielectron decay [see estimates for the corresponding widths in Eqs.~\eqref{eeX2}, \eqref{eewidth2}, and \eqref{eewidth0}].}
\label{fig:Xep}
\end{figure*}

In this section, we provide an estimate for the dielectron width of the $X(6200)$ and discuss its possible direct production in $e^+e^-$ annihilation. The corresponding diagrams are depicted in Fig.~\ref{fig:Xep}, where the employed $J/\psi\to e^+e^-$ vertex is defined to effectively describe the transition $J/\psi\to\gamma^*\to e^+e^-$. Since the $J/\psi$'s originating from the $X$ are highly nonrelativistic, we disregard their longitudinal components and arrive at convergent loop integrals. Then, assuming the quantum numbers of the $X$ to be $J^{PC}=2^{++}$, after a straightforward calculation\footnote{We employed the FeynCalc~10.1.0 package \cite{Mertig:1990an,Shtabovenko:2023idz} to simplify the analytical calculations and
LoopTools \cite{vanOldenborgh:1989wn} for a numerical evaluation of the scalar loop.} we arrive at the amplitude,
\be
\mathcal{A} = 2(2eg_{J/\psi})^2g_X\epsilon_{jk}\bar u (p_-)\gamma_j \slashed{I} \gamma_k u(-p_+),
\label{ampl}
\ee
where $u$'s stand for the electron and positron spinors and $I$ is the triangle loop integral,
\be
\slashed{I} = \frac{1}{i}\int \frac{d^4 p}{(2\pi)^4} G_e(p)G_{J/\psi}(p-p_-)G_{J/\psi}(p_++p),
\ee
with the propagators 
\begin{align}
&G_e(p)=\frac{\slashed{p}+m_e}{p^2-m_e^2+i\epsilon}\approx \frac{1}{\slashed{p}+i\epsilon},\nonumber\\[-2mm]
\label{Gs}\\[-2mm]
&G_{J/\psi}(p)=\frac{1}{p^2-m_{J/\psi}^2+i\epsilon},\nonumber
\end{align}
$\epsilon_{jk}=\epsilon_{kj}$ for the symmetric polarisation tensor of the $X_2(6200)$, and the overall factor 2 in the amplitude \eqref{ampl} accounts for the equivalent contributions from the two diagrams in Fig.~\ref{fig:Xep}. Setting $m_X\approx 2\mJ$, it is easy to calculate that
\be
I_\mu=-\frac{1}{64\pi^2 \mJ^2}\left(1-\mJ^2 C_0\right)(p_--p_+)_\mu%+{\cal O}(m_e^2)
,
\ee
with
\be
C_0\equiv C_0(0,0,4\mJ^2,\mJ^2,0,\mJ^2)
\approx-\frac{2.47}{\mJ^2}
\ee
for a standard scalar three-point loop function.

The effective coupling constant $2eg_{J/\psi}$ for the decay $J/\psi\to e^+e^-$ can be determined from the dielectron width $\Gamma_{J/\psi}^{ee} = \frac{4}{3}\alpha g_{J/\psi}^2 m_{J/\psi}$. Then the dielectron width of the tensor state $X_2(6200)$ reads
\be
\Gamma_{X_2}^{ee} =\frac{9g_X^2}{80\pi^3\mJ}\left(\frac{\Gamma_{J/\psi}^{ee}}{\mJ}\right)^2\left(1-\mJ^2 C_0
\right)^2,
\label{eeX2}
\ee
where the subscript $J=2$ is used to pinpoint the employed assumption on the spin of the $X(6200)$. The coupling $g_X$
can be found from the residue at the $X$ pole. To this end, we use that, near the $X$ pole, the $J/\psi$-$J/\psi$ scattering amplitude in Eq.~\eqref{eq:T} can be represented in the form
\be
T(s\simeq m_X^2)\approx\frac{g_X^2}{s-m_X^2}.
\ee

In Table~\ref{tab:GammaXee}, we collect the numerical values of $g_X$ evaluated for different combined fits to the data obtained above as well as the corresponding estimates for the dielectron width of the $X(6200)$ treated as a $2^{++}$ tensor. From this table, one can deduce an order-of-magnitude estimate,
\be
\Gamma_{X_2}^{ee} \simeq 0.01~\mbox{eV},
\label{eewidth2}
\ee
which appears somewhat smaller but may still be of the same order of magnitude typical for tensor charmonia. For example, the results for the generic ground-state tensor charmonium reported in Refs.~\cite{Kuhn:1979bb,Kivel:2015iea} read, respectively, 
\be
\Gamma_{\chi_{c2}(1P)}^{ee} =0.014~\mbox{eV (VMD)},~0.07~\mbox{eV (NRQCD)},
\label{chi2}
\ee
where in parentheses we quote the method used in the calculations, with VMD for vector meson dominance and NRQCD for nonrelativistic quantum chromodynamics. A similar numerical result was also obtained in Ref.~\cite{Shi:2023ntq} for the dielectron width of the $X_2(4014)$ as a molecular spin-two partner of the $X(3872)$.\footnote{Remarkably, a dielectron width of a similar magnitude was also found in Ref.~\cite{Denig:2014fha} for the $1^{++}$ exotic state $X(3872)$.}
Then, further following the logic of Ref.~\cite{Shi:2023ntq}, we can estimate 
the number of directly produced $X_2(6200)$ states at the planned Super $\tau$-Charm Facility \cite{Achasov:2023gey} at the level of $10^6$-$10^8$ per year, where we multiplied the expected 1-yr integrated luminosity around 1~ab$^{-1}$ by the $X_2$ production cross section evaluated as
\be
\sigma_C=\frac{20\pi\Gamma_{X_2}^{ee}}{\Gamma_{X}m_{X}^2}\sim 10^0\mbox{-}10^2~\mbox{pb},
\ee
for $\Gamma_{X}=0.1$-10~MeV.
The $X(6200)$ may also be detected in various open-charm final states such as $D\bar D$ or $\Lambda_c\bar \Lambda_c$. In this case, one $c\bar c$ pair annihilates into a gluon which then creates light quark-antiquark pairs, so the corresponding inclusive width could be roughly estimated as $\Gamma_{\eta_c}/\alpha_s(m_c)\sim 100$~MeV, with $\Gamma_{\eta_c}$ for the $\eta_c$ total width \cite{Anwar:2017toa}. Considering also the detection efficiency for each charmed hadron approximately at the level from a few per cent to about 10\%~\cite{ParticleDataGroup:2024cfk}, the number of the events of the $X(6200)$ open-charm decays (such as $D\bar D$, $\Lambda_c\bar \Lambda_c$, and so on) anticipated to be collected annually can be estimated as $10^3$-$10^5$.

On the other hand, in agreement with general expectations (see, for example, Ref.~\cite{Chanowitz:2005du}), the leading contribution to the $e^+e^-$ decay amplitude of a scalar state $X_0(6200)$ with $J^{PC}=0^{++}$ scales as $m_e$, as it originates from the numerator of the electron propagator $G_e(p)$ in Eq.~\eqref{Gs}. Consequently, the dielectron width $\Gamma_{X_0}^{ee}$ of a scalar state $X_0(6200)$ vanishes in the limit of a strictly massless lepton and remains
strongly suppressed when the lepton mass is finite,
\be
\Gamma_{X_0}^{ee} \sim \left(\frac{m_e}{\mJ}\right)^2 \Gamma_{X_2}^{ee},
\label{eewidth0}
\ee
where we used that the only typical mass scale in the loop is $m_{J/\psi}$.
As a result, this scalar state is not observable in direct production in $e^+e^-$ collisions even at the Super $\tau$-Charm Facility. Therefore, a potential direct production of the $X(6200)$ in the electron-positron annihilation can indirectly indicate its quantum numbers to be $2^{++}$.
Similar arguments also apply to higher fully charm tetraquarks that couple to di-$J/\psi$ under the molecular assumption.

\begin{table*}[t]
\caption{Estimates for the $X_2(6200)$ coupling to a $J/\psi$-$J/\psi$ pair (see also the colour coding in Fig.~\ref{fig:poles}) and its dielectron width.}
\label{tab:GammaXee}
\resizebox{\textwidth}{!}{
\begin{ruledtabular}
\begin{tabular}{ccccc}
&Two-channel bound&Two-channel virtual &Three-channel &Three-channel\\[-1mm]
&state & state or resonance & fit~1 & fit~2\\
\hline
$|g_X|$ [GeV] & [7.04,9.87]& [10.64,15.25]&[19.05,21.92] &[10.36,14.79]\\
$\Gamma^{ee} _{X_2 (6200)}$ [eV] & [0.002,0.004]& [0.005,0.010]& [0.016,0.022] & [0.005,0.010]\\
\end{tabular}
\end{ruledtabular}
}
\end{table*}

\section{Line shapes in complimentary channels}

\begin{figure}[t]
\includegraphics[width=0.95\linewidth]{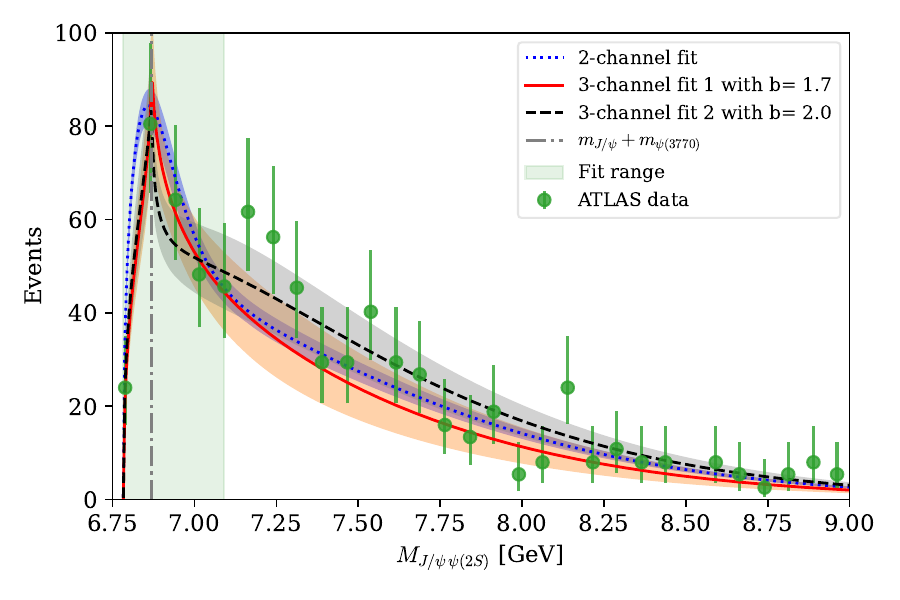}
\caption{The $J/\psi\psi(2S)$ invariant mass distribution obtained for the two- and three-channel models. The parameters of the scattering amplitude in Eq.~\eqref{eq:T} are fixed from the combined analysis of the data in the di-$J/\psi$ channel while the production-channel-related parameters $\alpha_3$ and $b$ are fitted to the ATLAS data from Ref.~\cite{ATLAS:2023bft} at $\sqrt{s}<7.1$ GeV (see Table~\ref{tab:psipsipb} for the parameter values). The dash-dotted vertical line indicates the position of the $J/\psi\psi(3770)$ threshold.}
\label{fig:Jpsipsip}
\end{figure}

Parameters of the scattering matrix $T(E)$ determined from the combined fits to the data, as explained in Sec.~\ref{sec:model} and collected in Tables~\ref{tab:params2c} and \ref{tab:params3c}, can further be used to study the line shapes in complimentary channels. In particular, in Fig.~\ref{fig:Jpsipsip}, we plot the line shapes in the channel $J/\psi\psi(2S)$ and compare the result with the data provided by the ATLAS Collaboration in Ref.~\cite{ATLAS:2023bft}. All the parameters of the scattering matrix are previously fully fixed as explained above, while the parameters $\alpha$ and $b$ are fitted to the ATLAS data to take into account the differences in the production mechanisms in the studied channels $J/\psi J/\psi$ and $J/\psi\psi(2S)$. A detailed description of the visible structures in the experimental signal would require an appropriate extension of the model validity range by including additional coupled channels, which lies beyond the scope of the present research. Meanwhile, we conclude on a decent overall description of the data with our theoretical curves even beyond the energy range of the formal validity of the employed coupled-channel models, $\sqrt{s}\lesssim 7.1$~GeV.

\begin{table*}[htb]
 \caption{The parameters in the $J/\psi\psi(2S)$ channel for the three-channel fit with the other parameters fixed to the central values in Table~\ref{tab:params3c}.
 }
 \label{tab:psipsipb}
 \resizebox{\textwidth}{!}{
 \begin{ruledtabular}
 \begin{tabular}{cccc}
 three-channel fit& $\alpha_3$ & $b$& $\chisq$ \\
 \hline
 fit 1 & $2441^{+657}_{-720}$& $1.71^{+0.30}_{-0.16}$& $0.08$\\
 %\hline
 fit 2 & $2072^{+597}_{-646}$& $2.04^{+0.38}_{-0.19}$& $0.29$ \\
 %\hline
 
 \end{tabular}
 \end{ruledtabular}
 }
\end{table*}

\section{Conclusions}

In this paper, we extend the approach of Ref.~\cite{Dong:2020nwy}, initially applied to the LHCb data on double-$J/\psi$ production in proton-proton collisions, to include recent data from the CMS and ATLAS Collaborations. A combined analysis of all three datasets is performed and the existence of the near-threshold state $X(6200)$ predicted in Ref.~\cite{Dong:2020nwy} is confirmed. We improve on the extraction of the parameters of this new state employing the combined fits to the data built for the two- and three-channel models and confirm that for most of the parameter space it has a molecular nature. The dielectron width of the $X(6200)$ evaluated under its plausible assignment as a $2^{++}$ di-$J/\psi$ molecule takes a reasonable value comparable with a similar estimate for $X(4014)$, also performed under the molecular assumption. Then, the corresponding estimates predict possible copious direct production of the tensor $X(6200)$ in $e^+e^-$ collisions at the planned Super $\tau$-Charm Facility. On the contrary, a scalar molecular $X(6200)$ has a tiny di-electron width and cannot be observed this way.

We further studied the complimentary $J/\psi\psi(2S)$ channel and analysed the ATLAS data in this final state. We concluded that, with the parameters of the multichannel scattering amplitude fully fixed from the combined analysis of the data in the di-$J/\psi$ channel, the ATLAS data could be well reproduced by fitting only two parameters specific for a particular production channel. Similar data from the LHCb and CMS Collaborations and, possibly, data for the $J/\psi\psi(3770)$ final state would potentially allow one to better fix the position of the pole describing the $X(6900)$ state. However, it may require an extension of the employed model to a larger number of coupled channels with the thresholds lying above 7.1~GeV. Meanwhile the number of parameters of such an extended model would considerably increase, and it remains to be seen whether or not the additional parameters can be reliably fixed from the data.

Some extra light on the existence and quantum numbers of the $X(6200)$ can be shed by employing complimentary theoretical and numerical approaches. In particular, in a recent paper \cite{Meng:2024czd} using methods of lattice QCD, the interaction in the channels $\eta_c\eta_c$ and $J/\psi J/\psi$ with the quantum numbers $J^{PC}=0^{++}$ and $2^{++}$, respectively, was found to be slightly repulsive, which is not consistent with the existence of near-threshold poles in these systems.
To further investigate the case, the second version of Ref.~\cite{Meng:2024czd} disentangles individual contributions from different Wick contractions. In particular, the authors find that both soft-gluon exchanges (denoted as contraction type ``D'' therein) that hadronise in the form of light-meson (pion and kaon) exchanges and the exchanges through heavy $c\bar{c}$ pairs (denoted as contraction type ``C'' therein) treated in a single-channel approximation individually lead to an attraction in the double-$J/\psi$ system. 
However, their combined effect, including interference, was surprisingly found to produce a weak repulsion. It therefore remains to be seen what mechanisms could allow one to override the $1/m_{c\bar{c}}^2$ suppression of the contribution from the $c\bar c$ exchanges (type C) compared with the soft gluons exchanges (type D). One might argue in favour of some $\mathcal{O}(N_c)$ enhancement (with $N_c$ for the number of colours) of the C-type contraction compared to the D-type one (see, for example, the analysis in Ref.~\cite{Guo:2013nja}).
However, this enhancement is unlikely to overcome the aforementioned suppression factor. 

It should also be noted that the interaction potential between two heavy quarkonia driven by the soft-gluon exchange mechanism 
strongly depends on the pion mass \cite{Dong:2021lkh}
(see also Refs.~\cite{Collins:2024sfi,Abolnikov:2024key} for the pion mass dependence of the pole position in the systems with charm). Thus, it remains to be seen whether or not the findings of Ref.~\cite{Meng:2024czd} stand a chiral extrapolation from the unphysically large pion mass employed in the calculations (from 300 to 365~MeV) to the physical point.

On the other hand, a support of the possible existence of a fully charm tetraquark with the mass around 6.2~GeV was provided in Ref.~\cite{Wang:2022xja} using the QCD sum rules method. The $0^{++}$ state is predicted to be slightly lighter than the $2^{++}$ one.

\begin{acknowledgements}
We would like to thank Yu Meng for helpful discussions regarding the lattice QCD results. 
This work is supported in part by the National Key R\&D Program of China under Grant No. 2023YFA1606703; by the Chinese Academy of Sciences (CAS) under Grants No.~YSBR-101 and No.~XDB34030000; and by the National Natural Science Foundation of China (NSFC) under Grants No. 12125507, No. 12361141819, and No. 12047503.
Work of A.N. was supported by Deutsche Forschungsgemeinschaft (Project No. 525056915). A.N. and C.H. also acknowledge the support from the CAS President’s International Fellowship Initiative (PIFI) (Grants No.~2024PVA0004 and No.~2025PD0087).
\end{acknowledgements}

%\bibliographystyle{apsrev4-2-fk}
%\bibliography{refs.bib}

%

\end{document}